\documentclass[prb,floatfix,twocolumn,showpacs,amsmath,amssymb]{revtex4}
\usepackage{graphicx}
\usepackage{dcolumn}
\usepackage{bm}

\begin{document}

\title{
Quantum oscillations of magnetization in the tight-binding electrons on honeycomb lattice 
}

\author
{Keita Kishigi}
\affiliation{Faculty of Education, Kumamoto University, Kurokami 2-40-1, 
Kumamoto, 860-8555, Japan}

\author{Yasumasa Hasegawa}
\affiliation{Department of Material Science, 
Graduate School of Material Science, 
University of Hyogo, Hyogo, 678-1297, Japan}

\date{\today}

\begin{abstract}
We show that the new quantum oscillations of the magnetization can occur when 
the Fermi surface consists of points (massless Dirac points) or even when 
the chemical potential is in a energy gap by studying the tight-binding electrons on a honeycomb lattice in a uniform magnetic field. The quantum oscillations of the magnetization as a function 
of the inverse magnetic field are known as the de Haas-van Alphen (dHvA) oscillations and the frequency is proportional to the area of the Fermi surface.
The dominant period of the new oscillations corresponds 
to the area of the first Brillouin zone and its phase is zero. 
The origin of the new 
quantum oscillations is the characteristic magnetic field dependence of 
the energy known as the Hofstadter butterfly and the
 Harper broadening of Landau levels. 
The new oscillations are not caused by the crossing of the chemical potential and Landau levels, which is the case in the dHvA oscillations. 
The new oscillations can be observed experimentally in systems with large supercell such as graphene antidot lattice or ultra cold atoms in optical lattice at an external magnetic field of a few Tesla when the area of the supercell is ten thousand times larger than that of graphene.
\end{abstract}

\date{\today}

\maketitle

\section{introduction}

The Shubnikov-de Haas (SdH) oscillations and the de Haas-van Alphen (dHvA) oscillations are 
powerful tools for observing the Fermi surface in conductors.\cite{Shoenberg84}
The magnetoresistance for the SdH and the magnetization for the dHvA oscillate periodically 
as a function of the inverse of a magnetic field ($H$), respectively. 
The extremal area of the Fermi surface in the plane perpendicular to the magnetic field is obtained from the period of the oscillations. 
In a semi-classical approximation the energy of electrons in the two-dimensional systems is quantized into Landau levels $\varepsilon_n$ due to a uniform magnetic field as 
\begin{equation}
 F(\varepsilon_n) = (n+\gamma) \frac{2 \pi e H}{\hbar c},
 \label{eqquantization}
\end{equation}
where $F(\varepsilon_n)$ is an area of the Fermi surface for $H=0$ in the wave-number space with the chemical potential $\mu=\varepsilon_n$, $n$ is Landau index, 
$e$ is the electron charge, $c$ is the speed of light, $\hbar$ is the Planck constant divided by $2 \pi$ and $\gamma$ is the phase factor. 
For the normal electrons, $\gamma=1/2$. For the massless Dirac fermions, $\gamma=0$, which are realized in graphene\cite{Novo2005} and $\alpha$-(BEDT-TTF)$_2$I$_3$.\cite{Tajima2013} 
For the normal electrons with effective mass $m$, Landau levels are quantized as
\begin{equation}
\varepsilon_n^{normal} =\frac{e \hbar}{m c} \left( n+\frac{1}{2}\right) H
\end{equation}
and the oscillatory part of the magnetization with the constant $\mu$ is given by 
the generalized Lifshitz and Kosevich (LK) formula
\cite{LK,Shoenberg84,champel,KH,Igor2004PRL,Igor2011,Sharapov},
\begin{equation}
M^{\rm LK} = - \frac{e}{2\pi^2 c\hbar} 
 \frac{F}{\frac{\partial F}{\partial \mu}}
\sum_{l=1}^{\infty}\frac{1}{l} R_{T}^{(l)}
\sin\left[2\pi l(\frac{f}{H}+\gamma) \right],
\label{LK_0}
\end{equation}
where
$R_T^{(l)} = \frac{\lambda l}{\sinh \lambda l}$
is the temperature reduction factor for the $l$-th harmonic,  
$\lambda =\frac{2 \pi^2 k_B T}{\hbar \omega_c}$,
$k_B$ is the Boltzmann constant, $T$ is the temperature and 
$\omega_c=eH/(cm)$ is the cyclotron frequency.
Due to the crossing of $\mu$ and Landau levels the magnetization oscillates 
periodically as a function of $1/H$ with the frequency 
\begin{equation}
f = \frac{c \hbar F}{2 \pi e}.
\label{eqfrequency}
\end{equation}
We have not taken account of the effects of Zeeman splitting and impurities.
The magnetization changes in a so-called inverse saw-tooth pattern as a function of $1/H$, since 
the coefficient of the $l$-th harmonics is proportional to $1/l$ at $T=0$.
When the number of electrons instead of the chemical potential is fixed, the chemical potential also oscillates and the saw-tooth pattern is inverted. 
Recently the LK formula is shown to be applicable for Dirac electrons 
in the case of small $\mu$ if we use the appropriate $n$ and $H$ dependences of Landau levels and cyclotron 
frequency.\cite{Igor2004PRL,Igor2011,Sharapov}

In the LK formula, however, the broadening of Landau levels due to the periodic potentials 
or the tight-binding nature of electrons is not taken into account. The broadening of Landau levels is known as the Harper broadening\cite{Hofstadter,Harper}, which makes the Hofstadter Butterfly diagram\cite{Hofstadter}. 
When the magnetic flux through the unit cell is $p/q$ times the flux quantum $\phi_0 = 2\pi \hbar c/e$, where $p$ and $q$ are mutually prime integers, Landau levels split into $p$ bands in the weak periodic potential case and the Bloch band splits into $q$ bands in the tight-binding model with one orbit in a unit cell and into $2q$ bands in the tight-binding model on a honeycomb lattice. Hall conductance is quantized when $\mu$ is in the $r$-th gap and it is given by the Chern number $t_r$ obtained from the Diophantine equation $r=q s_r  + p t_r $\cite{TKNN,HK2006,HK_2013}. 
The total energy of electrons is minimized when the magnetic field with one flux quantum per each electron is applied\cite{HLRW, HHKM}. Recently, the Hofstadter butterfly diagrams are observed experimentally in ultra cold atoms in optical lattices\cite{aidel,miyake} and moire superlattices\cite{Dean}.
In graphene anti-dot lattice\cite{Pedersen}, the energy band is obtained to be similar to the Hofstadter butterfly diagram. 

The magnetization in the tight-binding model has been studied by many authors\cite{M_1995,K_1995,sandhu,so,Gat_NJP,Gat,Mtaut,Gv_2007,Xu2008}. In the previous studies the oscillations of the magnetization are thought to be caused by the crossing of $\mu$ and Landau levels (i.e., the dHvA oscillations). 
\cite{M_1995,K_1995,sandhu,so,Gat_NJP,Gat,Mtaut,Gv_2007,Xu2008}. 
Gat and Avron\cite{Gat_NJP,Gat} have shown 
as a function of $\mu$ the magnetization near commensurate magnetic fluxes analytically in the semi-classical approximation in a square lattice. They have shown that besides the dHvA oscillations, which is zero when $\mu$ is at the center of the gap, the mean magnetization has the contributions of the Berry phase and the 
Wilkinson-Rammal (WR) phase\cite{Gat}.
As we will show below, the oscillations of magnetization exist even when $\mu$ is fixed in the middle of the gap (namely, $\mu$ does not cross Landau levels), where the quantized Hall conductance is zero. Thus it is not clear whether the new oscillations are caused by Berry or WR phases or not. Taut {\it et al}.\cite{Mtaut} reported rapid oscillations of the magnetization numerically in the tight-binding model on the square lattice, in addition to the dHvA oscillations, where there is always the Fermi surface at $H=0$. 
In this paper we study the total energy and the magnetization as a function of $H$ for $\mu=0$ in the tight-binding electrons on the honeycomb lattice.

\section{Model}

\begin{figure}[bt]
\begin{flushleft} \hspace{0.5cm}(a) \end{flushleft}\vspace{-0.0cm}
\includegraphics[width=0.53\textwidth]{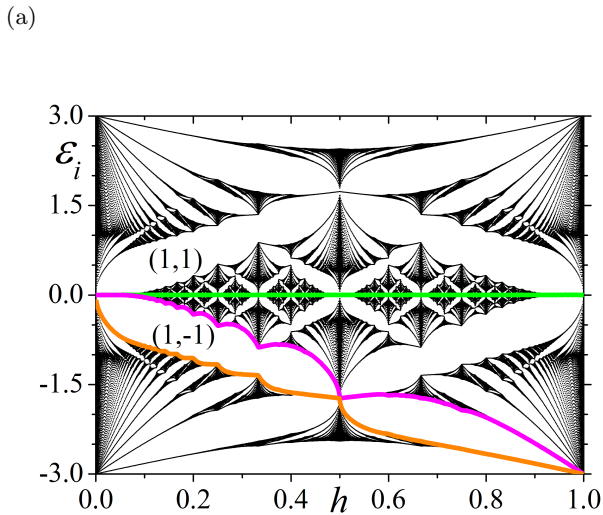}\vspace{-0.5cm}
\begin{flushleft} \hspace{0.5cm}(b) \end{flushleft}\vspace{-0.0cm}
\includegraphics[width=0.53\textwidth]{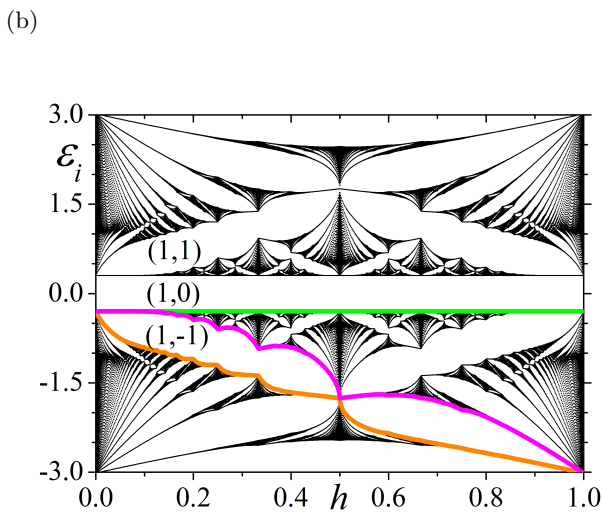}
\vspace{-0.5cm}
\caption{
Hofstadter butterfly diagrams of tight-binding electrons
on honeycomb lattice (a) and that with the difference between sublattices A and B ($\varepsilon_A=-\varepsilon_B=0.3$) (b).
In the half filled case the chemical potential is $\mu=0$ independent of $h$. 
In (b), $\mu$ is in the gap. (1,1), (1,0) and (1,-1) are the index of the gap ($s_r$,$t_r$).
}
\label{fig1ab}
\end{figure}

The Hamiltonian of the tight-binding electrons with nearest-neighbor 
hoppings in the magnetic field
 is given by
\begin{equation}
\mathcal{H} = -\sum_{(i,j)} t_{i,j} e^{i\theta_{ij}}c_i^{\dagger} c_j 
  + \varepsilon_A \sum_{i \in A} c_i^{\dagger} c_i 
  + \varepsilon_B \sum_{i \in B} c_i^{\dagger} c_i, 
\end{equation}
where
\begin{equation}
\theta_{ij} =  \frac{2 \pi}{\phi_0} \int_{\mathbf{r}_i}^{\mathbf{r}_j} \mathbf{A} \cdot d \boldsymbol{\ell}
\end{equation}
and $\mathbf{A}$ is the vector potential. 
The hopping integrals $t_{i,j}$ between site $\mathbf{r}_i$ and $\mathbf{r}_j$ are $t=1$, when 
$\mathbf{r}_i$ and $\mathbf{r}_j$ are the nearest sites 
and otherwise $t_{i,j}=0$.
We have introduced the site energies $\varepsilon_A$ and $\varepsilon_B$ for A and B sublattices. When $H=0$ and $\varepsilon_A=\varepsilon_B$, the Fermi surface in the half-filled case consists of two Dirac points. When the inversion symmetry is broken by the different potential at each sublattice, a finite gap is opened. 
We study the cases that the magnetic field $\mathbf{H} = \nabla \times \mathbf{A}$ 
is uniform and the flux through the unit cell $\phi=\sqrt{3}Ha^2/2$, where $a$ is a lattice constant, is taken as a rational number,
\begin{equation}
\frac{\phi}{\phi_0}=\frac{p}{q} \equiv h. 
\end{equation}
Hereafter, we use $h$ instead of $H$. 
The energy is obtained as the
eigenvalues $\varepsilon_{i,{\bf k}}$ of $2 q \times 2 q$ matrix for each wave number $\mathbf{k}=(k_x, k_y)$. 
Thermodynamic potential per site $(\Omega)$, the total energy per site at $T=0$ ($E$) and 
the magnetization at $T=0$ ($M$) are calculated by 
\begin{equation}
\Omega=-\frac{k_BT}{2qN} \sum_{i=1}^{2q}\sum_{{\bf k}} \log\left
\{\exp\left(\frac{\mu-\varepsilon_{i,{\bf k}}}{k_BT} \right) +1
\right\},
\end{equation}
\begin{equation}
E=\frac{1}{2qN}\sum_{\varepsilon_{i,{\bf k}}\leq\mu}  (\varepsilon_{i,{\bf k}}-\mu)
\end{equation}
and 
\begin{equation}
M=-\frac{\partial E}{\partial h}, 
\end{equation}
respectively, where $N$ is the number of points of ${\bf k}$. 
Since the eigenvalues depend on $\mathbf{k}$, 
there are $2q$ energy bands. This property is different from the semi-classical quantization, where Landau levels are treated
as delta functions.  
For sufficiently large $q$ (e.g. $q=907$), however, the width of each band is narrow and we can represent the energy for each band at fixed $\mathbf{k}$ (e.g. $\mathbf{k}=0$).
Therefore, $N=1$ and we obtain the energy of each band
as $\varepsilon_i$ ($i=1, 2, \cdots, 2q$) as a function of $h$.    
Then we obtain the Hofstadter butterfly diagram as shown in Fig.~\ref{fig1ab}.

\section{Results and discussion}

\begin{figure}[bt]
\includegraphics[width=0.56\textwidth]{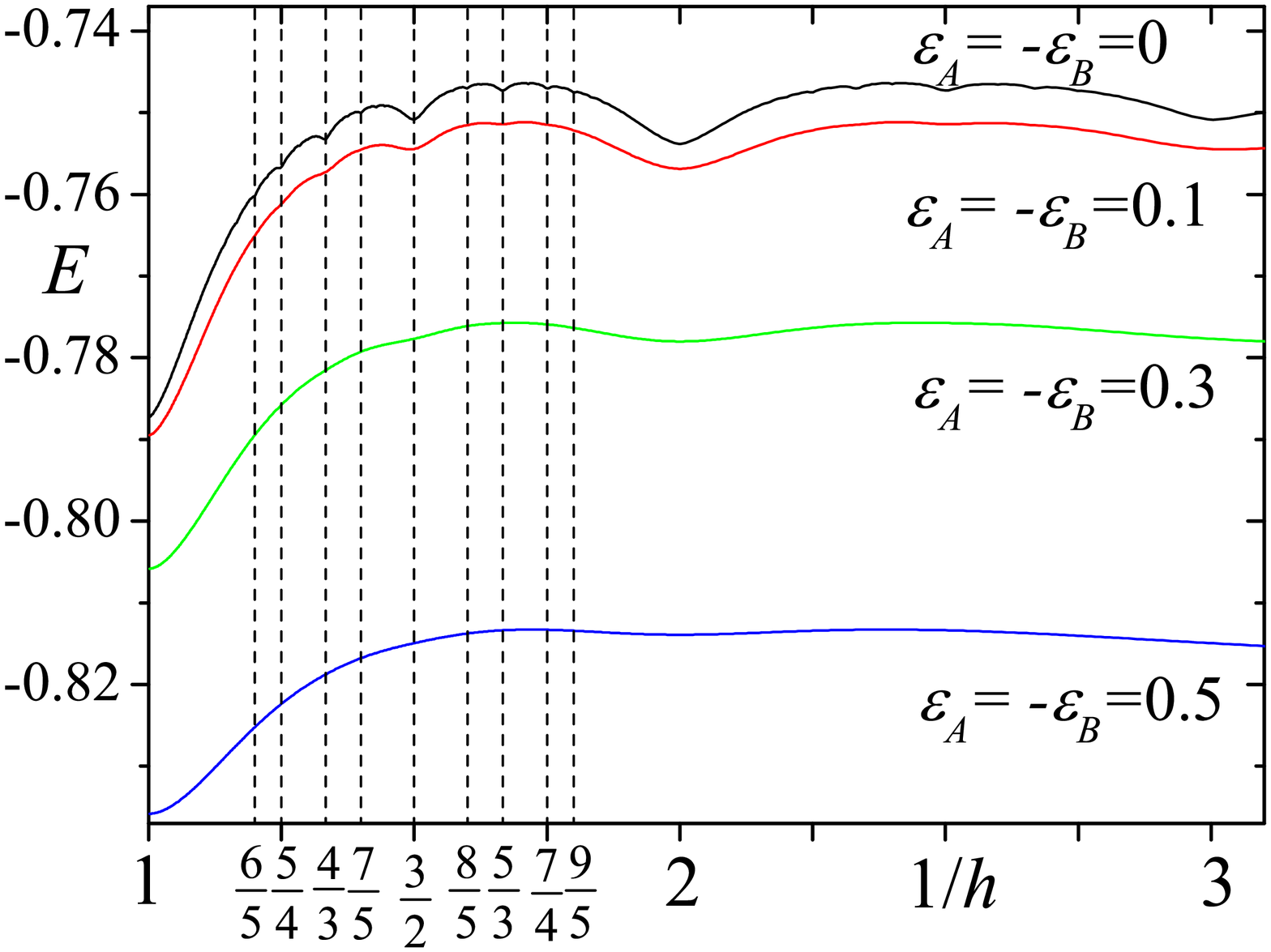}
\caption{Total energies as a function of $1/h$ for some on-site potentials. 
}
\label{fig2d}
\end{figure}

\begin{figure}[bt]
\begin{flushleft} \hspace{0.5cm}(a) \end{flushleft}\vspace{-0.5cm}
\includegraphics[width=0.53\textwidth]{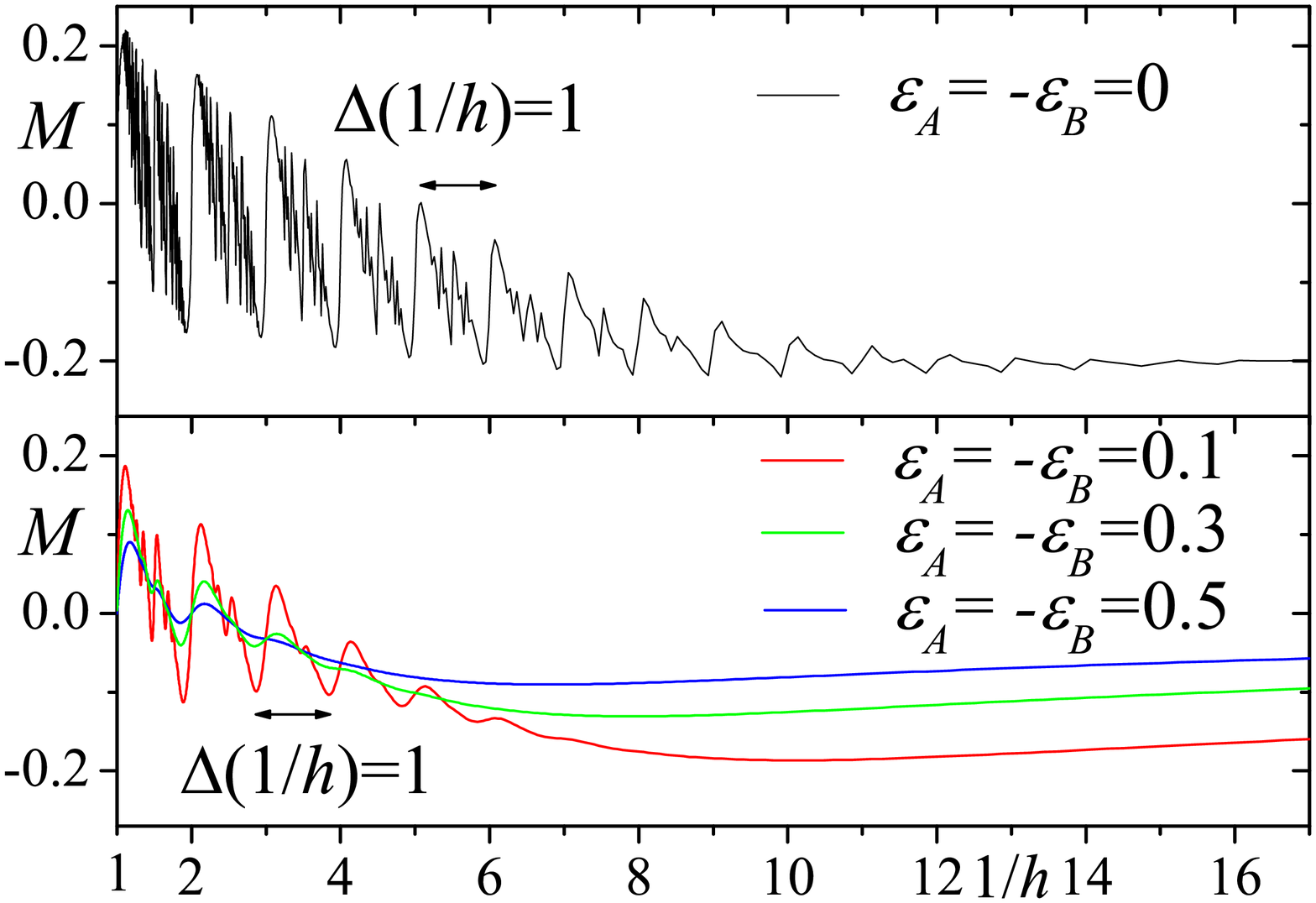}\vspace{-0.5cm}
\begin{flushleft} \hspace{0.5cm}(b) \end{flushleft}\vspace{-0.5cm}
\includegraphics[width=0.53\textwidth]{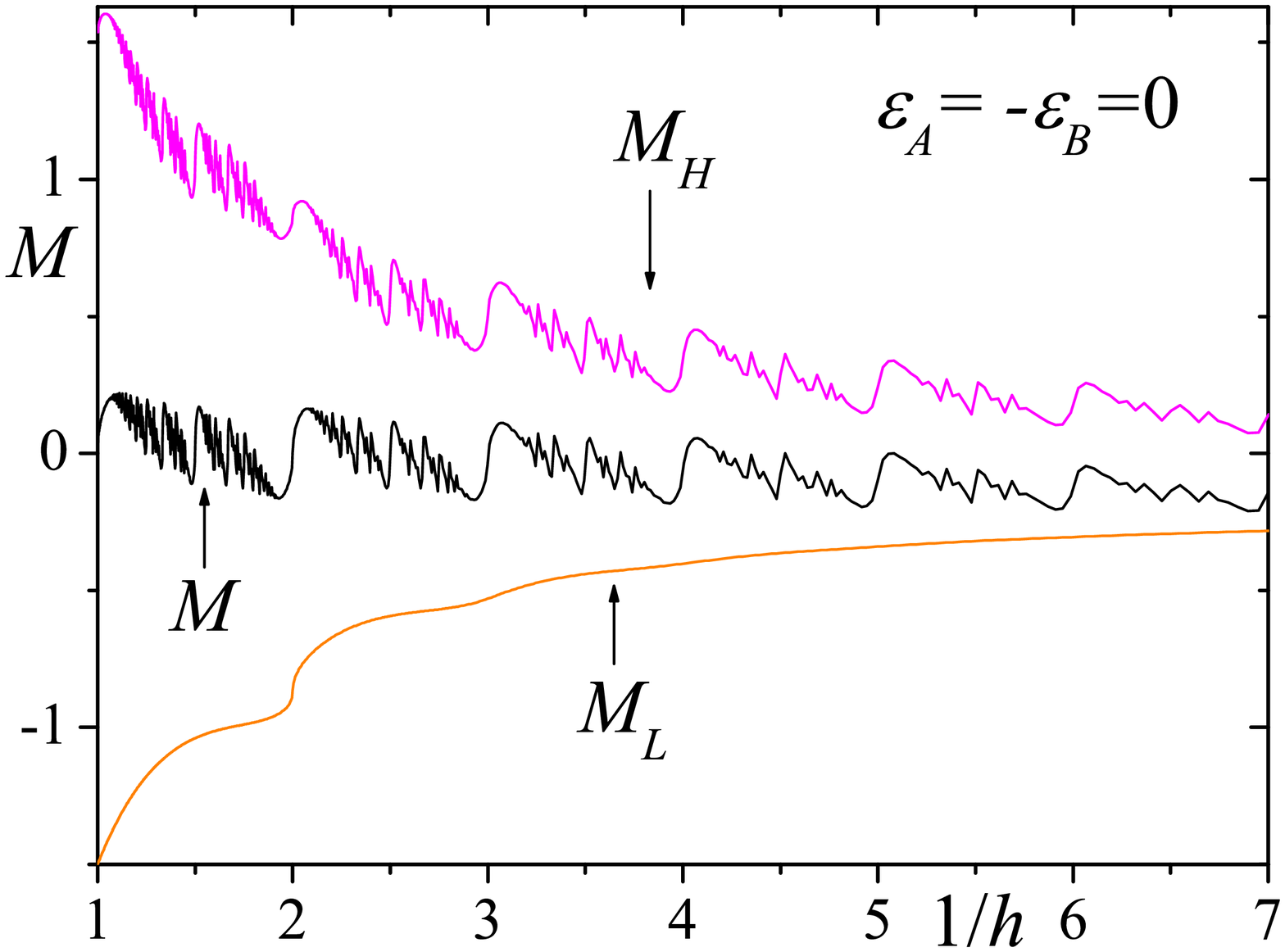}
\caption{Magnetization of half-filled case of tight-binding electrons on honeycomb lattice as a function of $h$.
In (b) contributions from $n=0$ band ($M_H$) and others ($M_L$) are plotted separatedly.
}
\label{figfig3}
\end{figure}

The total energy is a continuous function of $1/h$ but it has many dips as shown in Fig. \ref{fig2d}, in which we plot 
total energies for the half-filled ($\mu=0$) case as a function of $1/h$ for several values of $\varepsilon_A=-\varepsilon_B$. 
We calculate the magnetization by the numerical differentiation by changing $p$ with fixed $q=907$. The essentially same results are obtained even if we change the value of $q$, (for example, $q=499$, $467$, etc). Therefore, the numerical errors 
due to the differentiation instead of derivative is expected to be negligibly small. 
The magnetization at $T=0$ as a function of $1/h$ is shown in Fig.~\ref{figfig3} for $\varepsilon_A =-\varepsilon_B=0$, $0.1$, $0.3$ and $0.5$.

\begin{figure}[bt]
\centering
\includegraphics[width=0.53\textwidth]{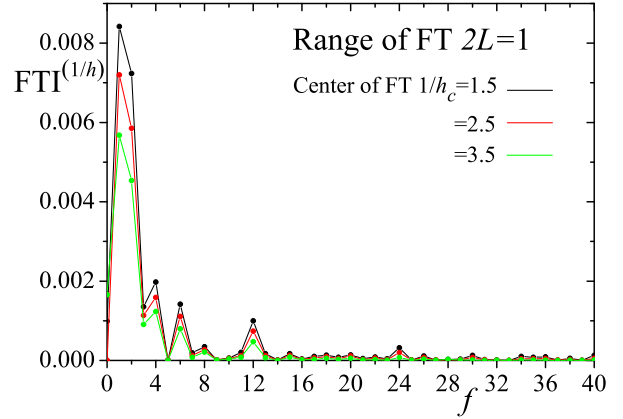}
\caption{FTIs for $\varepsilon_A=-\varepsilon_B=0$ of Fig. \ref{figfig3} (a).}
\label{figfig1g}
\end{figure}

In the case of $\varepsilon_A=-\varepsilon_B=0$, 
the Fermi surface at $h=0$ is two Dirac points. 
When $\varepsilon_A = -\varepsilon_B \neq 0$, there is the energy gap 
between two bands at $h=0$ and the chemical potential is in the gap,
i.e. there is no Fermi surface.  
In the generalized LK formula [Eq. (\ref{LK_0})], the magnetization oscillates with the frequency $f$, which is proportional to the 
area of the Fermi surface [Eq.~(\ref{eqfrequency})]. Therefore, the oscillations of the magnetization are not expected in the half-filled 
tight-binding electrons on the honeycomb lattice. 
However, as shown in Fig.~\ref{figfig3} (a) the magnetization oscillates, 
although the oscillations of the magnetization on $1/h$ are not perfectly periodic. The amplitudes of oscillations decrease as $1/h$ increases. The shape is not a perfect saw-tooth but it is similar to the saw-tooth pattern for fixed electron numbers rather than the inverse saw-tooth pattern for the fixed chemical potential, although it is chaotic for small $1/h$. 
The most dominant period of the oscillations 
of $M$ as a function of $1/h$ is 1, which corresponds to the area of the first Brillouin zone in the LK formula. The phase of the oscillations corresponds to zero ($\gamma=0$) in the LK formula. 
Since the magnetization is not a perfect periodic function of $1/h$, we calculate the
Fourier transform, choosing the center $1/h_c$ and a finite range $2L$, as
\begin{equation}
  \mathrm{FTI}^{(1/h)}(f, \frac{1}{h_c},L) =\left|\frac{1}{2L} 
  \int_{\frac{1}{h_c}-L}^{\frac{1}{h_c}+L} M (h) e^{2\pi i \frac{f}{h}} d \left(\frac{1}{h}\right)\right|^2.
\end{equation}
Since we perform the Fourier transform in the finite range $2L$, we take $f=j/(2L)$ with integer $j$.
We plot the Fourier transform intensities (FTIs) for $\varepsilon_A=-\varepsilon_B=0$, 
$2L=1$ and $1/h_c=1.5$, $2.5$ and $3.5$ in Fig.~\ref{figfig1g}. 
If the oscillations would have a perfect saw-tooth pattern, FTIs would decrease as $1/f^2$ and not depend 
on the choice of $1/h_c$.  As shown in Fig.~\ref{figfig1g}, the components of $f=1, 2$, $4$, $6$, $12$, $24$ etc. are large and these of
$f=5, 7$, $9$ etc. are small. Amplitudes become small when we take 
larger $1/h_c$, but the $f$-dependences are similar. 

\begin{figure}[bt]
\includegraphics[width=0.53\textwidth]{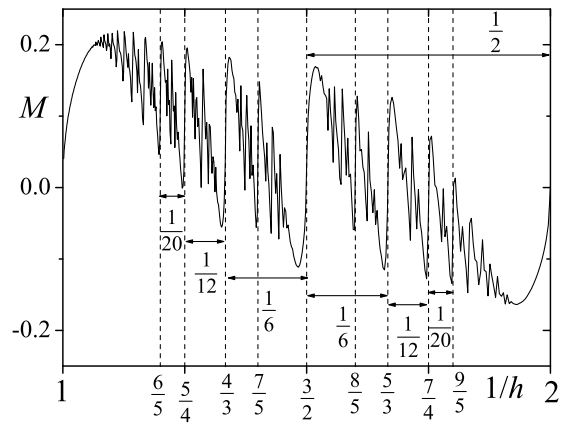}
\caption{The close-up figure of the Fig.~\ref{figfig3} (a) for $\varepsilon_A=-\varepsilon_B=0$ (magnetization of half-filled case of tight-binding electrons on honeycomb lattice as a function of $1/h$). 
}
\label{figfig1c}
\end{figure}

In order to see the magnetization as a function of $1/h$ in detail, we plot $M$ for $1 \leq 1/h \leq 2$
in Fig.~\ref{figfig1c}.
There are many jump-like structures when $1/h$ is a rational number ($1/h=q/p$) written by small integers ($p$ and $q$). 
However, it is clearly seen that they are not discontinuous jumps but continuous sharp cliffs at $1/h=1$, $4/3$, $3/2$, $5/3$, $2$, 
etc. There are the large saw-tooth-like oscillations with the periods of 1, 
$\frac{1}{2}, \frac{1}{6}$ and $\frac{1}{12}$, which contribute to the peaks of FTI$^{(1/h)} (f, 1/h_c, L)$ for $f=1,2,4,6,12$ and 24.


In order to make clear the origin of the magnetic-field dependence 
of the magnetization, we calculate the contributions of the
different part of Hofstadter butterfly diagram separately.
We define $E_H$, $E_L$, $M_H$ and $M_L$ as
\begin{align}
E_H &= D \sum_{\varepsilon_n \in R_H} \varepsilon_n, \\
E_L &= D \sum_{\varepsilon_n \in R_L} \varepsilon_n, 
\end{align}
\begin{align}
M_H&= - \frac{\partial E_H}{\partial h}
\end{align}
and
\begin{align}
M_L&= - \frac{\partial E_L}{\partial h},
\end{align}
where $R_H$ is the set of eigenstates with the energy between $\varepsilon_i=0$ and the large gap starting from $\varepsilon_i=0$ at $h=0$ 
[states between green and pink curves in Fig.~\ref{fig1ab} (a)], $R_L$ is the set of eigenstates with the energy below the large gap 
[states below the orange curve in Fig.~\ref{fig1ab} (a)].
In Fig.~\ref{figfig3}~(b) we plot $M_H$, $M_L$ and $M=M_H+M_L$ as functions of $1/h$ for $\varepsilon_A=-\varepsilon_B=0$. 
The oscillatory dependence of the magnetization comes from $M_H$. 
The states between green and pink curves in Fig.~\ref{fig1ab} (a) are the $n=0$ Landau level in the continuum limit. 
Namely, the oscillations of $M_H$ in Fig.~\ref{figfig3}~(b) are thought to be caused by the broadening of Landau levels due to
the tight-binding electrons. 



\begin{figure}[bt]
\includegraphics[width=0.53\textwidth]{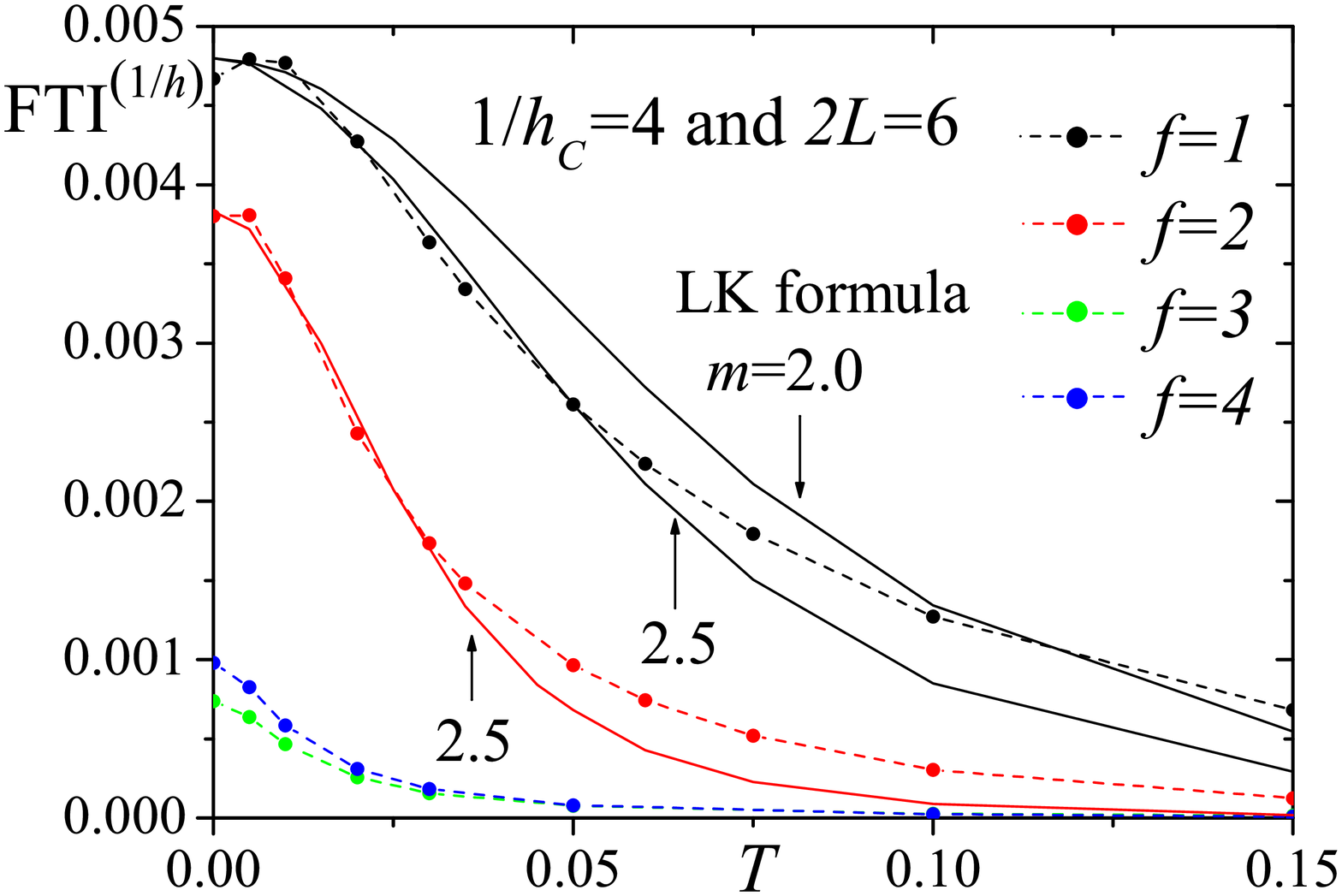}
\caption{$T$-dependences of the FTIs with $f=1, 2, 3, 4$. 
The Fourier transformation is done 
in the region $1\leq 1/h\leq 7$. 
Solid lines are $T$-dependences of the FTIs of LK formula with $m=2.0$ and $m=2.5$.  
The numerically calculated cyclotron mass at $\mu=-0.8$ in the tight-binding model is about 2.5. 
}
\label{fig:5}
\end{figure}

Next, we study the temperature dependence of the FTIs,
which are shown in Fig.~\ref{fig:5}. 
Although the LK formula cannot be applied in the present case 
($R_T^{(l)}$is always $1$, if we set the effective mass to be zero), we try to fit the temperature reduction as
the reduction factor for normal electrons with the effective mass.  We obtain that
the effect of the temperature can be fitted by the effective mass $m=2.5$, where the unit of the mass is $\hbar^2/(ta^2)$, at low temperature region 
($T \lesssim 0.04$), where the unit of $T$ is $t$, but it deviates as the temperature becomes large.

\section{Conclusion}

We have shown that the oscillations of the magnetization as a function of the magnetic field exist even when the area of the Fermi surface vanishes, if the tight-binding electrons on the honeycomb lattice are studied. 
Since the chemical potential stays in the energy gap when $\epsilon_A \neq \epsilon_B$, the new quantum oscillations are not caused by the crossing of the chemical potential and the energy bands (Landau levels), which is the origin of the dHvA oscillations in the case of semi-classical approximation. 
They come from the complex structure of the $n=0$ Landau level in the Hofstadter butterfly diagram. The origin of the new oscillations can also be considered as the simultaneous crossings of Landau levels ($\pm n$ starting from the bottom and top of the energy ($\epsilon = \mp 3$) at $h=0$), which do not change
the energy gap at $\epsilon =0$. The effect of temperature reduces the amplitudes of the oscillations, 
but it is not described by the temperature reduction factor in the LK formula. 
We can fit the temperature dependence by the temperature reduction factor, $R_{T}^{(l)}$, only in small region of temperature.

It is difficult to observe the new quantum oscillations of magnetization experimentally, because $h\simeq 1/16$ means $H\simeq 5000$ T from $a\simeq 0.246$ nm of graphene and $H\simeq 250$ T from the area of unit cell\cite{Bender} of $\alpha$-(BEDT-TTF)$_2$I$_3$\cite{Tajima2013}, respectively. 
However, if the area of the supercell in graphene antidot lattices\cite{Pedersen} or ultracold atoms on the optical lattice\cite{nature2012} is taken to be about ten thousand times larger than that of graphene, the new quantum oscillations of the magnetization 
may be observed at a few Tesla. In this paper, we have neglected the Zeeman energy. 
When the energy bands for up and down spins are overlapped by the Zeeman splitting, the new quantum oscillations will be suppressed. 
However, it will be possible to make a system have an energy gap 
($|\epsilon_A-\epsilon_B|$) at $H=0$ larger than $\mu_B H$, where $\mu_B$ is the Bohr magneton and $H$ is a few Tesla. 
In that system the chemical potential stays in the energy gap even when the Zeeman energy is taken into account, and the new oscillations are possible to be observed.


\begin{thebibliography}{99}


\bibitem{Shoenberg84}
D.~Schoenberg, {\it Magnetic oscillations in metals} 
(Cambridge University Press: Cambridge, 1984).




\bibitem{Novo2005}
K. S. Novoselov, A. K. Geim, S. V. Morozov, D. Jiang, M. I. Katsnelson, I. V. Grigorieva, S. V. Dubonos, and A. A. Firsov, 
Nature {\bf 438}, 197 (2005).

\bibitem{Tajima2013}
N. Tajima, T. Yamauchi, T. Yamaguchi, M. Suda, Y. Kawasugi, H. M. Yamamoto, 
R. Kato, Y. Nishio, and K. Kajita, Phys. Rev. B 88, 075315 (2013).



\bibitem{LK}
I. M. Lifshitz and A. M. Kosevich, Sov. Phys. JETP {\bf 2} 636 (1956).

\bibitem{champel}
T. Champel, Phys. Rev. B{\bf 64}, 054407 (2001).

\bibitem{KH}
K. Kishigi and Y. Hasegawa, Phys. Rev. B {\bf 65}, 205405, (2002).


\bibitem{Igor2004PRL}
I. A. Luk'yanchuk and Y. Kopelevich, Phys. Rev. Lett. {\bf 93} 
166402 (2004).

\bibitem{Igor2011}
I. A. Luk'yanchuk, Low Temperature Physics {\bf 37}, 45 (2011).

\bibitem{Sharapov}
S. G. Sharapov, V. P. Gusynin, and H. Beck, Phys. Rev. {\bf B69} (2004) 075104.


\bibitem{Hofstadter} 
D. R. Hofstadter, Phys. Rev. B {\bf 14}, 2239 (1976).



\bibitem{Harper} 
P. G. Harper, Proc. Phys. Soc. Lond. A {\bf 68}, 874 (1955).



\bibitem{TKNN}
D. J. Thouless, M. Kohmoto, M. P. Nightingale, and M. den Nijs, Phys. Rev. Lett. {\bf 49}, 405 (1982).

\bibitem{HK2006}
Y. Hasegawa and M. Kohmoto, 
Phys. Rev. B {\bf 74}, 155415 (2006). 


\bibitem{HK_2013}
Y. Hasegawa and M. Kohmoto, Phys. Rev. B {\bf 88}, 125426 (2013).


\bibitem{HLRW}
Y. Hasegawa, P. Lederer, T. M. Rice and P. B. Wiegmann, 
Phys. Rev. Lett. {\bf 63}, 907 (1989). 


\bibitem{HHKM}
Y. Hasegawa, Y. Hatsugai, M. Kohmoto and G. Montambaux,
Phys. Rev. B {\bf 41}, 9174 (1990). 




































































\bibitem{aidel}
M. Aidelsburger, M. Atala, M. Lohse, J. T. Barreiro, B. Paredes, and I. Bloch, Phys. Rev. Lett. {\bf 111}, 185301 (2013).

\bibitem{miyake}
H. Miyake, G.A. Siviloglou, C.J. Kennedy, W.C. Burton, and 
W. Ketterle, Phys. Rev. Lett. {\bf 111}, 185302 (2013).


\bibitem{Dean}
C. R. Dean, L. Wang, P. Maher, C. Forsythe, F. Ghahari, Y. Gao, J. Katoch, M. Ishigami, 
P. Moon, M. Koshino,
T. Taniguchi, K.Watanabe, K. L. Shepard, J.Hone, and P. Kim, Nature {\bf 497}, 598 
(2013). 

\bibitem{Pedersen} 
J. G. Pedersen and T. G. Pedersen, 
Phys. Rev. B {\bf 87}, 235404 (2013).


\bibitem{M_1995}
K. Machida, K. Kishigi and Y. Hori, Phys. Rev. {\bf B 51}, 8946 (1995). 


\bibitem{K_1995}
K. Kishigi, M. Nakano, K. Machida, and Y. Hori, 
J. Phys. Soc. Jpn. {\bf 64}, 3043 (1995). 

\bibitem{sandhu}
P. S. Sandhu, J. H. Kim, and J. S. Brooks, Phys Rev. B{\bf 56} 11566 (1997).


\bibitem{so}
S. Y. Han, J. S. Brooks and J. H. Kim Phys. Rev. Lett. {\bf 85}, 1500 (2000).





\bibitem{Gat_NJP}
O. Gat and J. E. Avron, New J. Phys. {\bf 5}, 44.1-44.8 (2003).

\bibitem{Gat}
O. Gat and J. E. Avron, Phys. Rev. Lett., {\bf 91}, 186801 (2003).

\bibitem{Mtaut}
M. Taut, H. Eschrig and M. Richter, Phys. Rev. B {\bf 72}, 165304 (2005).

\bibitem{Gv_2007}
V. M. Gvozdikov and M. Taut, Phys. Rev. B {\bf 75}, 155436 (2007).

\bibitem{Xu2008}
W. H. Xu, L.P. Yang, M. P. Qin and T. Xiang, Phys. Rev. B {\bf 78}, 241102 (2008). 







\bibitem{nature2012} 
L. Tarruell, D. Greif, T. Uehlinger, G. Jotzu, T. Esslinger, and J. L. McChesney, 
Nature {\bf 483}, 305 (2012).


\bibitem{Bender} 
K. Bender, I. Hennig, D. Schweitzer, K. Dietz, H. Endres and 
H. J. Keller, Mol. Cryst. Liq. Cryst. {\bf 108}, 359 (1984). 


\end{thebibliography}
\end{document}